\documentstyle[preprint,aps,epsfig,floats]{revtex}

\renewcommand{\d}{\rm{d}}
\newcommand{\kv}{{\bf k}_{\perp}}
\newcommand{\lv}{{\bf l}_{\perp}}
\newcommand{\qv}{{\bf q}_{\perp}}
\newcommand{\bv}{{\bf b}_{\perp}}
\newcommand{\ac}{{\centerline{\bf Acknowledgments}}}
\tightenlines
\draft
\begin{document}
\date{\today}
\preprint{\vbox{\hbox{RUB-TPII-17/98}\hbox{PNU-NTG-04/98}}}
\title{Infrared-finite factorization and renormalization scheme
       for exclusive processes. Application to pion form factors\\
       }
\author{N. G. Stefanis,${}^{1}$\thanks{Email:
        stefanis@tp2.ruhr-uni-bochum.de}
        W. Schroers,${}^{2}$\thanks{Email:
        wolfram@theorie.physik.uni-wuppertal.de}
        and
        H.-Ch. Kim ${}^{3}$\thanks{Email:
        hchkim@hyowon.pusan.ac.kr}
        }
\address{${}^{1}$ Institut f\"ur Theoretische Physik II,   \\
                  Ruhr-Universit\"at Bochum,               \\
                  D-44780 Bochum, Germany                  \\
                  [0.3cm]
         ${}^{2}$ Fachbereich Physik,                      \\
                  Universit\"at Wuppertal,                 \\
                  D-42097 Wuppertal, Germany               \\
                  [0.3cm]
         ${}^{1}$ Department of Physics,                   \\
                  Pusan National University,               \\
                  Pusan 609-735, Republic of Korea         \\
         }
\maketitle
\newpage
\begin{abstract}
We develop and discuss an infrared-finite factorization and
optimized renormalization scheme for calculating exclusive
processes which enables the inclusion of transverse degrees of
freedom without entailing suppression of calculated observables,
like form factors. This is achieved by employing an analytic,
i.e., infrared stable, effective coupling $\alpha _{s}(Q^{2})$
which removes the Landau singularity at
$Q^{2}=\Lambda _{\rm QCD}^{2}$ by a power-behaved correction. The
ensuing contributions to the cusp anomalous dimension, related to
the Sudakov form factor, and to the quark anomalous dimension,
which controls evolution, lead to enhancement of the hard part of
exclusive amplitudes, calculated in perturbative QCD. The
phenomenological implications of this framework are analyzed
by applying it to the pion's electromagnetic form factor and the
pion-photon transition.
\end{abstract}
\pacs{11.10.Hi, 12.38.Bx, 12.38.Cy, 13.40.Hq, 13.40.Gp}
%Keywords: Renormalization group evolution
%          Perturbative calculations
%          Summation of perturbation theory
%          Sudakov effects
%          Renormalons
%          Hadron wave functions
%          Electromagnetic meson decays
%          Electromagnetic form factors

\newpage

\section{I\lowercase{ntroduction}}
\label{sec:intro}

In the last few years, several works
\cite{Zak92,DMW96,SS97,AZ97a,Gru97}
(among many others) addressed the possibility of power corrections to
the strong running coupling, beyond the operator product expansion.
Such corrections, which are subleading in the ultraviolet (UV) region,
correspond to nonanalytical contributions to the $\beta$-function.

The existence of these power corrections, if proven true, would
greatly affect our understanding of nonperturbative QCD effects.
For instance, a power correction to $\alpha _{\rm s}$ gives rise to a
linear term in the interquark static potential \cite{AZ97b}.
On a more speculative level, one may argue \cite{AZ97a} that the source
of such terms are small-size fluctuations in the nonperturbative QCD
vacuum, perhaps related to magnetic monopoles in dual QCD or nonlocal
condensates.
Besides, and in practice, a power-behaved contribution at low scales
can be used to remove the Landau singularity, present in perturbation
theory, supplying in this way an infrared (IR) stable effective coupling
\cite{DMW96,SS97,Gru97}.

The aim of the present work is to develop in detail a factorization and
renormalization scheme, which self-consistently incorporates such a
nonperturbative power correction, and then use it to assess and
explore exclusive processes. We do not, however, propose to involve
ourselves in the discussion of whether or not such power corrections
have a foundamental justification.
We consider the ambiguity in removing the Landau pole as resembling the
ambiguity in adopting a particular (non-IR-finite) renormalization
scheme in perturbative QCD. The justification for such a procedure will
be supplied a posteriori by the self-consistent incorporation of
higher-order perturbative corrections and by removing any IR-sensitivity
of calculated hadronic observables.

A key ingredient of our approach is that the modified effective
coupling will be taken into account not only in the factorized
short-distance part, i.e., through the fixed-order perturbation
expansion, but also in the resummed perturbative expression for
soft-gluon emission and in the renormalization-group controlled
evolution of the factorized parts.

To this end, we adopt as a concrete power-corrected effective coupling,
an analytic model for $\alpha _{\rm s}$, proposed by Shirkov and
Solovtsov \cite{SS97}, which yields an IR-finite and universal effective
(running) coupling.
This model combines Lehmann analyticity with the renormalization group
to remove the Landau singularity at $Q^{2}=\Lambda _{\rm QCD}^{2}$,
without employing adjustable parameters, just by modifying the
logarithmic behavior of $\alpha _{\rm s}$ by a simple (nonperturbative)
power term (minimality of the model).

At the present stage of evidence, it would be, however, premature to
exclude other parametrizations, since neither the sign of the power
correction nor the size of its exponent are established, and one could
introduce further modifications \cite{AZ97a,Ale98}.

Continuing our previous exploratory study \cite{SSK98}, we further
extend and test our theoretical framework by also including into the
calculation of the pion form factor the next-to-leading order (NLO)
perturbative contribution to the hard scattering amplitude (see, e.g.,
\cite{MNP98} and earlier references cited therein).

The ultimate goal of the present analysis is not to obtain results in
perfect agreement with the data, but to expose and discuss the
conceptual advantages of our scheme relative to previous conventional
approaches \cite{LS92,JK93,TL98}.

The major advantage of such a theoretical framework, the object of this
paper, is that it enables the inclusion of transverse degrees of
freedom, primordial (i.e., intrinsic) \cite{JK93} and those originating
from (soft) gluonic radiative corrections \cite{LS92}, without entailing
suppression of perturbatively calculated observables, viz., the pion
form factor.
This enhancement is due to power-term generated contributions to the
anomalous dimensions of the cusped Wilson line, related to the Sudakov
form factor, and such to the quark wave function which governs
evolution.

Although most of our considerations refer to the pion as a case study
for the proposed framework, the reasoning can be extended to describe
three-quark systems as well. This will be reported elsewhere.

The outline of the paper is as follows. In the next section we briefly
discuss the essential features of the IR-finite effective coupling.
In sect. \ref{sec:IRF} we develop and present our theoretical scheme.
Sect. \ref{sec:piffNLO} extends the method to the next-to-leading
order contribution to the hard-scattering amplitude.
In sect. \ref{sec:valid} we discuss the numerical analysis of the
electromagnetic pion form factor, revolving around the appropriate
kinematic cuts in the evaluation of the IR-modified Sudakov form factor
which comprises additional nonleading contributions. We also provide
arguments for the appropriate choice of the renormalization scale.
In this section we also show the theoretical prediction for the
pion-gamma transition form factor, derived with our theoretical
framework, as an independent justification of the approach.
Finally, in sect. \ref{sec:sum}, we summarize our results and draw our
conclusions.

\section{M\lowercase{odel for} QCD \lowercase{effective coupling}}
\label{sec:alpha_s}

The key element of the analytic approach of Shirkov and Solovtsov is
that it combines a dispersion-relation approach, based on local duality,
with the renormalization group (RG) to bridge the regions of small and
large momenta, providing universality at low scales. The approach is an
extension to QCD of a method originally formulated for QED \cite{Red58}.

At the one-loop level, the ghost singularity is removed by a simple
power correction and the effective coupling reads
\begin{equation}
  \bar{\alpha}_{\rm s}^{(1)}(Q^{2})
=
  \frac{4\pi}{\beta _{0}}
  \left[
          \frac{1}{\ln \left( Q^{2}/\Lambda ^{2} \right)}
        + \frac{\Lambda ^{2}}{\Lambda ^{2} - Q^{2}}
  \right] \; ,
\label{eq:oneloopalpha_an}
\end{equation}
%Eq (1) Analytic model for running QCD coupling constant; 1-loop level
where $\Lambda \equiv \Lambda _{\rm QCD}$ is the QCD scale parameter.

This model has the following interesting properties. It provides a
nonperturbative regularization at low scales and leads to a universal
value of the coupling constant at zero momentum
$
   \bar{\alpha}_{\rm s}^{(1)}(Q^{2}=0)
 = 4 \pi /\beta _{0}
\simeq
   1.396
$
(for three flavors), defined only by group constants.
No adjustable parameters are involved and no implicit ``freezing'',
i.e., no saturation hypothesis of the coupling constant is invoked.

Note that this limiting value (i) does not depend on the scale parameter
$\Lambda$ -- this being a consequence of RG invariance -- and (ii)
extends to the two-loop order, i.e.,
$
 \bar {\alpha}_{\rm s}^{(2)}(Q^{2}=0)
=
 \bar {\alpha}_{\rm s}^{(1)}(Q^{2}=0)
\equiv
 \bar {\alpha}_{\rm s}(Q^{2}=0)
$.
(In the following the bar is dropped.)
Hence, in contrast to standard perturbation theory, the IR limit of
the coupling constant is stable, i.e., does not depend on higher-order
corrections and is therefore universal. As a result, the running
coupling constant also shows IR stability. This is tightly connected
to the nonperturbative contribution
$\propto \exp (-4\pi /\alpha \beta _{0})$
which ensures analytic behavior in the IR domain by eliminating the
ghost pole at $Q^{2} = \Lambda ^{2}$, and extends to higher loop orders.
Besides, the stability in the UV domain is not changed relative to the
conventional approach, and UV perturbation theory is preserved.

At very low-momentum values, say, below 1~GeV, $\Lambda _{\rm QCD}$
in this model deviates from that used in minimal subtraction schemes.
However, since we are primarily interested in a region of momenta which
is much larger than this scale, the role of this renormalization-scheme
dependence is only marginal. In our investigation we use
$
 \Lambda _{\rm QCD}^{{\rm an} (\rm {n_{\rm f}}=3)}
=
 242~{\rm MeV}
$
which corresponds to
$
 \Lambda _{\rm QCD}^{{\overline{\rm MS}} (\rm {n_{\rm f}}=3)}
=
 200~{\rm MeV}
$.

The extension of the model to two-loop level is possible, though
the corresponding expression is too complicated to be given explicitly
\cite{SS97}. An approximated formula with an inaccuracy less than
$0.5\%$ in the region
$2.5 \, \Lambda < Q < 3.5 \, \Lambda$,
and practically coinciding with the exact result for larger values of
momenta, is provided by \cite{SS97}
\begin{equation}
  \alpha _{\rm s}^{(2)}(Q^{2})
=
  \frac{4\pi}{\beta _{0}}
  \left[
         \frac{1}{\ln \frac{Q^{2}}{\Lambda ^{2}}
       + \frac{\beta _{1}}{\beta _{0}^{2}}\,
         \ln \left(
                   1 + \frac{\beta _{0}^{2}}{\beta _{1}}
                       \ln \frac{Q^{2}}{\Lambda ^{2}}
            \right)}
      + \frac{1}{2}\, \frac{1}{1-\frac{Q^{2}}{\Lambda ^{2}}}
      - \frac{\Lambda ^{2}}{Q^{2}}D_{1}
  \right] \; ,
\label{eq:twoloopalpha_an}
\end{equation}
%Eq (2) Approximate expression for two-loop effective coupling
where
$
 \beta _{0}
=
 11 - \frac{2}{3}n_{\rm f}
=
 9
$,
$
 \beta _{1}
=
 102 - \frac{38}{3}n_{\rm f}
=
 64
$,
and $D_{1}=0.035$ for $n_{\rm f}=3$.

With experimental data at relatively low momentum-transfer values for
most exclusive processes, reliable theoretical predictions based on
perturbation theory are difficult to obtain. Both the unphysical Landau
pole of $\alpha _{\rm s}$ and IR instability of the factorized
short-distance part are affecting such calculations. It is precisely
for these two reasons that the Shirkov-Solovtsov analytic approach to
the QCD effective coupling can be profitably used for computing
amplitudes describing exclusive processes, like form factors,
\cite{LB80,CZ77,ER80}.
The improvements are then:
(i)   First and foremost, the nonperturbatively generated power
      correction modifies the Sudakov form factor
      \cite{Col89,BS89,KR87,Kor89,GKKS97} via the cusp anomalous
      dimension \cite{Pol79}, and changes also the evolution behavior of
      the soft and hard parts through the modified anomalous dimension
      of the quark wave function. This additional contribution to the
      cusp anomalous dimension is the source of the observed IR
      enhancement and helps taking into account {\it nonperturbative
      corrections in the perturbative domain}, thus improving the
      quality of the predictions.
(ii)  Factorization is ensured without invoking the additional
      assumption of ``freezing'' the coupling strength in the IR regime
      by introducing, for example, an (external) effective gluon mass to
      saturate color forces at large distances.
(iii) The Sudakov form factor does not have to serve as an IR protector
      against $\alpha _{\rm s}$ singularities. Hence the extra
      constraint of
      using the maximum between the longitudinal and the transverse
      scale, as argument of $\alpha _{\rm s}$, proposed in \cite{LS92}
      and used in subsequent works, becomes superfluous.
(iv)  The factorization and renormalization scheme we propose on that
      basis enables the optimization of the (arbitrary) constants which
      define the factorization and renormalization scales
      \cite{Col89,BS89,CS81,DS84}. This becomes important when including
      higher-order perturbative corrections.

\section{I\lowercase{nfrared-finite factorization and renormalization
         scheme}}
\label{sec:IRF}

Application of perturbative QCD is based on factorization, i.e., how
a short-distance part can be isolated from the large-distance physics.
But in order that observables calculated with perturbation theory are
reliable, one must deal with basic problems, like the resummation of
``soft'' logarithms, IR sensitivity, and the factorization and
renormalization scheme dependence.

It is one of the purposes of the present work to give a general and
thorough investigation of such questions.

The object of our study is the electromagnetic pion's form factor in the
spacelike region, which can be expressed as the overlap of the
corresponding wave functions between the initial and final pion states:
\cite{DY69}
\begin{equation}
  F_{\pi}\left(Q^{2}\right)
=
  e_{\rm q}
  \int_{0}^{1} {\d}x
  {\d}^{2}\kv
  \psi _{\pi}^{\rm out}\left( x, \lv \right)
  \psi _{\pi}^{\rm in} \left( x, \kv \right) \; ,
\label{eq:DYW}
\end{equation}
%Eq (3) Pion form factor as direct overlap of wave functions
%       (DYW formula)
where we have assumed dominance of the valence $q\bar q$ state,
with $e_{\rm q}$ denoting the charge of the struck quark, and where
\begin{equation}
  \lv
=
  \left\{\begin{array}{ll}
        \kv +(1-x)\qv \; ,
  \;\;\;\;\;\;\;\;\;\; & {\rm struck} \;\, {\rm quark} \\
        \kv - x\qv
  \;\;\;\;\;\;\;\;\;\; & {\rm spectators} \; .
\end{array}
\right.
\label{eq:transmom}
\end{equation}
%Eq (4) Transverse momenta of outgoing quarks
The wave function $\psi _{\pi} (x, \kv )$ is the amplitude for finding
a parton in the valence state with longitudinal momentum fraction $x$
and transverse momentum $\kv$.

In order to apply a hard-scattering analysis, we dissect the pion wave
function into a soft and a hard part with respect to a factorization
scale $\mu_{\rm F}$, separating the perturbative from the
nonperturbative regime, and write (in the light-cone gauge $A^{+}=0$)
\begin{equation}
    \psi _{\pi}\left( x, \kv \right)
 =
    \psi _{\pi}^{\rm soft}\left( x, \kv \right)
    \theta \left( \mu _{\rm F}^{2} - \kv ^{2} \right)
  + \psi _{\pi}^{\rm hard}\left( x, \kv \right)
    \theta \left( \kv ^{2} - \mu _{\rm F}^{2} \right) \; .
\label{eq:psidec}
\end{equation}
%Eq (5) Decomposition of pion wave function into soft and hard part
Then the large $k_{\perp}$ tail can be extracted from the soft wave
function via a single-gluon exchange kernel, encoded in the hard
scattering amplitude $T_{\rm H}$, so that \cite{LB80}
\begin{equation}
  \psi _{\rm hard}\left( x, \kv \right)
=
  \int_{0}^{1} {\d}y \int_{}^{}{\d}^{2}\lv
  T_{\rm H} \left( x, y, \lv ^{2};
  \alpha _{s} \! \left( \lv ^{2} \right)
            \right)
  \psi _{\rm soft}\! \left( y, \lv\right) \; .
\label{eq:hardgluonpot}
\end{equation}
%Eq (6) Extraction of hard tail of wave function via 1-gluon exchange
%       kernel
As a result, the pion form factor in Eq. (\ref{eq:DYW}) is expressed in
the factorized form
\begin{eqnarray}
\everymath{\displaystyle}
  F_{\pi} (Q^{2})
= &&
   \psi _{\rm soft}^{\rm out}
   \otimes
   \psi _{\rm soft}^{\rm in}
 + \psi _{\rm soft}^{\rm out}
   \otimes
   \left[ T_{\rm H} \otimes
          \psi _{\rm soft}^{\rm in}
   \right]
 +
   \left[ \psi _{\rm soft}^{\rm out} \otimes
          T_{\rm H}
   \right]
   \otimes
   \psi _{\rm soft}^{\rm in}
\nonumber \\
&& +
     \left[ \psi _{\rm soft}^{\rm out} \otimes
            T_{\rm H}
     \right]
     \otimes
     \left[ T_{\rm H} \otimes
            \psi _{\rm soft}^{\rm in}
     \right] + \ldots \; ,
\label{eq:softhard}
\end{eqnarray}
%Eq (7) Dissection of pion form factor in soft and hard part
where the symbol $\otimes$ denotes convolution defined by
Eq. (\ref{eq:hardgluonpot}).
The first term in this expansion is the soft contribution to the form
factor, with support in the low-momentum domain, that is not computable
with perturbative methods. The second term represents the leading order
(LO) hard contribution due to one-gluon exchange, whereas the last one
gives the NLO correction. We will not attempt to calculate the first
term here, but adopt for simplicity the result obtained by Kroll
and coworkers in \cite{JKR96}. For other, more sophisticated, attempts
to model the soft contribution to $F_{\pi}(Q^{2})$, we refer to
\cite{Rad84,Szc98}.

\begin{figure}
\tighten
\[
\psfig{figure=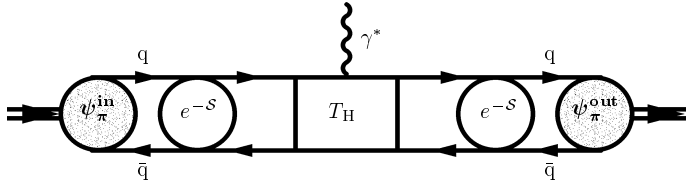,%
       bbllx=136pt,bblly=688pt,bburx=485pt,bbury=753pt,%
       width=10.5cm,silent=}
\]
\vspace{1cm}
\caption[fig:feynmangraph]
        {\tenrm Illustration of the factorized pion form factor,
         exhibiting the different regimes of dynamics. The wiggly line
         denotes the off-shell photon. Gluon exchanges are not
         explicitly displayed. The region of hard-gluon rescattering
         (LO and NLO) is contained in the short-distance part, termed
         $T_{\rm H}$. The blobs ${\rm e}(-S)$ represent in axial
         gauge Sudakov-type contributions, whereas nonperturbative
         effects are absorbed into the pion wave functions
         $\psi _{\pi}^{\rm in}$ and $\psi _{\pi}^{\rm out}$.
\label{fig:feypi}}
\end{figure}

We now employ a modified factorization prescription \cite{LS92,JK93},
which explicitly retains transverse degrees of freedom, and define
(see for illustration fig. \ref{fig:feypi})
\begin{equation}
  \psi _{\pi}^{\rm hard}
=
  \psi _{\pi}^{\rm soft}
  \left(
        \kv ^{2}\leq \frac{C_{3}^{2}}{b^{2}}
  \right)
  \exp \left[ - S \left( \frac{C_{1}^{2}}{b^{2}}
                         \leq \kv ^{2}
                         \leq C_{2}^{2}\xi ^{2}Q^{2}
                  \right)
       \right]
  T_{\rm Hard}\left( Q^{2} \geq \kv ^{2}
                     \geq \frac{C_{3}^{2}}{b^{2}}
              \right) \; ,
\label{eq:modpsi}
\end{equation}
%Eq (8) Modified factorization of pion wave function
with $b$, the variable conjugate to $k_{\perp}$, being the transverse
distance between the quark and the antiquark in the pion valence
Fock state. The Sudakov-type form factor $\exp (-S)$ comprises leading
and next-to-leading logarithmic corrections, arising from soft and
collinear gluons, and resums all large logarithms in the region where
$\kv ^{2}\ll Q^{2}$ \cite{CS81,DS84,DDT80}.
The presence of these logarithms results from the incomplete
cancellation between soft-gluon bremsstrahlung and radiative
corrections.
It goes without saying that the function $S$ includes
anomalous-dimension contributions to match the change in the running
coupling in a commensurate way with the changes of the renormalization
scale (see below for more details).

%-----------------------------------------------------------------------
%                            F I G U R E  1
%                          \label{fig:feypi}
%-----------------------------------------------------------------------
%
%---------------------------------------------------------------------

Going over to the transverse configuration space, the pion form factor
reads \cite{LS92}
\begin{eqnarray}
\everymath{\displaystyle}
  F_{\pi}\left(Q^{2}\right)
= &&
  \int_{0}^{1} {\d}x {\d}y \int_{-\infty}^{\infty}
                     \frac{{\d}^{2}\bv}{(4\pi )^{2}} \,
          {\cal P}_{\pi}^{\rm out}
          \left( y, b, P^{\prime}; C_{1}, C_{2}, C_{4}
          \right)
  T_{\rm H}\left(
                 x, y, b, Q; C_{3}, C_{4}
           \right)
\nonumber \\
&&
\times\;  {\cal P}_{\pi}^{\rm in}
             \left( x, b, P; C_{1}, C_{2}, C_{4}
             \right) \; ,
\label{eq:piffbspace}
\end{eqnarray}
%Eq (9) Pion form factor in transverse configuration space
where the modified pion wave function is defined in terms of matrix
elements, viz.,
\begin{eqnarray}
  {\cal P}_{\pi}
                \left( x, b, P, \mu
                \right)
& = &
  \int_{}^{|\kv |<\mu} {\d}^{2}\kv {\rm e}^{- i \kv \cdot \bv}
  {\widetilde{\cal P}}_{\pi}\left( x, \kv , P
       \right)
\nonumber \\
& = &
  \int_{}^{} \frac{{\d}z^{-}}{2\pi}
  {\rm e}^{-ix P^{+}z^{-}}
  {\left\langle 0 \left\vert
  {\rm T} \left(
                \bar{q}(0)\gamma ^{+}\gamma _{5}
                q\left(0,z^{-},\bv \right)
          \right)
  \right\vert \pi (P) \right\rangle}_{A^{+}=0}
\label{eq:matrel}
\end{eqnarray}
%Eq (10) Definition of modified pion wave function through matrix
%        elements evaluated on the light cone
with $P^{+}=Q/\sqrt{2}=P^{-\prime}$, $Q^{2}=-(P^{\prime}-P)^{2}$,
whereas the dependence on the renormalization scale $\mu$ on the rhs
of Eq.~(\ref{eq:matrel}) enters through the normalization scale of the
current operator evaluated on the light cone.
(Note that we set all light quark masses equal to zero and work in the
chiral limit, i.e., $M_{\pi}=0$.)

A few comments on the scales involved:
\begin{itemize}
\item The scale $C_{3}/b$ serves to separate perturbative from
      nonperturbative transverse distances (lower factorization scale).
      We assume that there exist some characteristic scale
      $
       b_{\rm nonp}^{-1} \simeq
       \langle \kv ^{2}\rangle ^{1/2}/x(1-x) \simeq 0.5
      $~GeV,
      related to the typical virtuality (offshellness) of vacuum
      quarks. This scale should also provide the natural starting point
      for the evolution of the pion wave function.
      In the following, we match the nonperturbative scale $C_{3}/b$
      with the scale $C_{1}/b$, where the resummation of soft gluons
      starts, i.e., we set $C_{1}=C_{3}$.
      The lower boundary of the scale $C_{1}/b$ is set by
      $\Lambda _{\rm QCD}$, though the results are not very sensitive
      to using a somewhat larger momentum scale, as we shall see later.
\item The resummation range in the Sudakov form factor is limited from
      above by the scale $C_{2}\xi Q$ (upper factorization scale).
      (Note that the constant $C_{2}$ here differs in notation by a
      factor of $\sqrt{2}$ relative to that used by Collins, Soper, and
      Sterman \cite{CS81}, i.e., $C_{2}^{\rm CSS}=\sqrt{2}C_{2}$.)
      This scale may be thought of as being an UV-cutoff for the (soft)
      Sudakov form factor, and enables this way a RG-controlled scale
      dependence governed by appropriate anomalous dimensions within
      this subsector of the full theory.
\item Analogously to these factorization scales, characterized by the
      constants $C_{1}, C_{2}$, and $C_{3}$, we have introduced an
      additional arbitrary constant $C_{4}$ to define the
      renormalization scale $C_{4}f(x,y)Q=\mu _{\rm R}$, which appears
      in the argument of the effective coupling $\alpha _{s}^{\rm an}$.
      The effective coupling plays a dual role: it describes the
      strength of the interaction at short distances, and controls via
      the anomalous quark dimension the lower and upper boundaries,
      respectively, for the evolution of $T_{\rm H}$, and
      ${\cal P}_{\pi}$ to the renormalization scale.
\end{itemize}
The appropriate choice of the unphysical and arbitrary constants $C_{i}$
will be discussed in our numerical analysis in sect. \ref{sec:valid}.

The ambiguities parametrized by the scheme constants $C_{i}$ emerge from
the truncation of the perturbative series and would be absent if one
would be able to derive all-order expressions in the coupling constant.
In fact, the {\it calculated} (pion) form factor depends implicitly on
both scales: the adopted renormalization scale via $\alpha _{s}$, and
the particular factorization scheme through the anomalous dimensions.
Since the latter also depend on $\alpha _{s}$, the factorization-scheme
and the renormalization-scheme dependences are correlated.
On the other hand, the {\it physical} form factor is independent of such
artificial scales, and satisfies
$
 \mu \frac{d F_{\pi}^{\rm phys} \left( Q^{2} \right)}{d\mu}
=
 0
$,
for $\mu$ being any internal scale.
Obviously, both scheme dependences should be treated simultaneously
and be minimized in order to improve the self-consistency of the
perturbative treatment. In order to render the perturbative prediction
reliable, the parameters $C_{i}$ should be adjusted in such a way, as
to minimize the influence of higher-order corrections, thus resolving
the scheme ambiguity. However, in the present investigation we are not
going to explicitly match the fixed-order NLO contributions with the
corresponding terms in the resummed expression for the ``soft''
logarithms. This check will be included in future work.

In Eq. (\ref{eq:piffbspace}), $T_{\rm H}$ is the amplitude for a quark
and an antiquark to scatter collinearly via a series of hard-gluon
exchanges, and, in LO perturbative expansion, it is given by
\begin{equation}
  T_{\rm H}\left(
                 x, y, b, Q; \mu _{\rm R}
           \right)
=
  8 C_{\rm F} \alpha _{\rm s}^{{\rm an}}(\mu _{\rm R}^{2})
  K_{0} \left( \sqrt{xy}\, bQ\right) \; .
\label{eq:T_Hbspace}
\end{equation}
%Eq (11) Hard-scattering amplitude for pion form factor in transverse
%        configuration space
This result is related to the more familiar momentum-space expression
\begin{equation}
  T_{\rm H}
        \left(
              x, y, \kv ,\lv ,Q, \mu _{\rm R}
        \right)
=
    \frac{16 \pi \, C_{\rm F} \,
    \alpha _{\rm s}(\mu _{\rm R}^{2})}{xyQ^{2}
  + \left( \kv + \lv \right)^{2}}
\label{eq:T_Hkspace}
\end{equation}
%Eq (12) Hard scattering amplitude for pion form factor depending on
%        transverse momenta
via the Fourier transformation
\begin{equation}
  T_{\rm H}
           \left(x, y, \kv , \lv , Q, \mu _{\rm R}
           \right)
=
  \int_{-\infty}^{\infty} {\d}{}^{2}\bv \,
  T_{\rm H}\left( x, y, b, Q, \mu _{R} \right)
  \exp \left[
             i \bv \cdot \left(
                                \kv + \lv
                         \right)
       \right] \; ,
\label{eq:fourierT_H}
\end{equation}
%Eq (13) Fourier transformation of hard scattering amplitude T_H
where use of the symmetry of $\psi _{\pi}$ under
$x \leftrightarrow 1-x \equiv \bar{x}$ has been made, and where
$C_{\rm F}=(N_{\rm c}^{2} - 1)/2N_{\rm c} = 4/3$ for $SU(3)$.

The amplitude
\begin{eqnarray}
\everymath{\displaystyle}
  {\cal P}_{\pi}\left( x, b, P\simeq Q, C_{1}, C_{2}, \mu
                \right)
=
 \exp \Biggl[
            && -  s\left( x, b, Q, C_{1}, C_{2} \right)
            -     s\left( \bar{x}, b, Q, C_{1}, C_{2} \right)
\nonumber \\
&&
            - 2 \int_{C_{1}/b}^{\mu} \frac{{\d}\bar\mu}{\bar\mu}\,
                \gamma _{q}\left(\alpha _{\rm s}^{\rm an}(\bar\mu )
                           \right)
      \Biggr]
  {\cal P}_{\pi}\left( x, b, C_{1}/b \right)
\label{eq:piamplbspace}
\end{eqnarray}
%Eq (14) Modified pion distribution amplitude including Sudakov
%        corrections and primordial k_perp distribution
describes the distribution of longitudinal momentum fractions of the
q\=q pair, taking into account the intrinsic transverse size of the
pion state \cite{JK93}, and comprising corrections due to soft real and
virtual gluons \cite{LS92}, including evolution to the renormalization
point.

The pion distribution amplitude evaluated at the factorization scale is
approximately given by
\begin{equation}
  {\cal P}_{\pi}\left( x, b, C_{1}/b \right)
\simeq
  \phi _{\pi}\left( x, C_{1}/b \right) \Sigma (x, b) \; .
\label{eq:inpiampl}
\end{equation}
%Eq (15) Pion distribution amplitude at factorization scale
In the present work, we follow Jakob and Kroll \cite{JK93} and
parametrize the distribution in the intrinsic transverse momentum
$\kv$ (or equivalently the interquark transverse distance $b$) in
the form of a nonfactorizing in the variables $x$ and $\kv$
(or $x$ and $b$) Gaussian function which is normalized to unity,
\begin{equation}
  \Sigma ( x, b )
=
  4\pi \, \exp \left[ - \frac{b^{2}}{4 g(x) \beta ^{2}} \right] \; ,
\label{eq:sigma_b}
\end{equation}
%Eq (16) Gaussian in b space for pion
where $g(x)=1/x\bar{x}$, with an appropriate width $\beta$ to be
specified below.

Neglecting transverse momenta in Eq. (\ref{eq:T_Hkspace}) (collinear
approximation), the only dependence on $k_{\perp}$ resides in the wave
function. Limiting the maximum value of $k_{\perp}$, these degrees of
freedom can be integrated out independently for the initial and final
pion states to give way to the corresponding pion distribution
amplitudes, which depend only implicitly on the cutoff momentum:
\begin{equation}
  \frac{f_{\pi}}{2\sqrt{2N_{\rm c}}} \,
  \phi _{\pi}\left( x, \mu ^{2} \right)
=
  \int_{}^{\mu ^{2}} \frac{{\d}^{2}\kv}{16 \pi ^{3}} \,
  \Psi _{\pi} ( x, \kv ) \; ,
\label{eq:pida}
\end{equation}
%Eq (17) Pion distribution amplitude
where
$f_{\pi} = 130.7$~MeV and $N_{\rm c}=3$.
Integrating on both sides of this equation over $x$ normalizes
$\phi _{\pi}$
to unity, i.e.,
$
 \int_{0}^{1}{\d}x \, \phi _{\pi}\left( x, \mu ^{2} \right)
=
 1
$
because the rhs is fixed to
$
 \frac{f_{\pi}}{2\sqrt{2N_{\rm c}}} \,
$
by the leptonic decay $\pi \rightarrow \mu ^{+} \nu _{\mu}$ for any
factorization scale.

Hence, the full (model) wave function for the pion takes finally the
form \cite{JK93}
\begin{equation}
  \Psi _{\pi}\left(x,\kv\right)
=
  \frac{16 \pi ^{2} f_{\pi}}{2\sqrt{2N_{\rm c}}}\,
  \phi (x)\, \beta ^{2} g(x)
  \exp\left[ - g(x) \beta ^{2}\kv ^{2} \right] \; .
\label{eq:piwfjk}
\end{equation}
%Eq (18) Pion wave function according to Jakob and Kroll

Let us now return to Eq. (\ref{eq:piamplbspace}).
The Sudakov form factor
$F_{\rm S}\left(\xi , b, Q, C_{1}, C_{2}\right)$,
i.e., the exponential factor in front of the wave function, will be
expressed as the expectation value of an open Wilson line along a
contour of finite extent, $C$, which follows the bent quark line in
the hard-scattering process from the segment with direction $P$ to that
with direction $P^{\prime}$ after being abruptly derailed by the hard
interaction which creates a ``cusp'' in $C$, and is to be evaluated
within the range of momenta termed ``soft'', confined within the range
limited by $C_{1}/b$ and $C_{2}\xi Q$
(where $\xi = x, \bar{x}, y, \bar{y}$).
(This means that the region of hard interaction of the Wilson line with
the off-shell photon is factorized out.)
Thus we have \cite{Col89,KR87,Kor89,GKKS97,KR92}
\begin{equation}
  F_{\rm S}\left(W(C)\right)
=
  \left\langle {\rm P}
  \exp \left(
             i g \int_{C}^{} dz
             \cdot t^{a}
             A^{a}(z)
       \right)
  \right \rangle _{\rm soft} \; ,
\label{eq:wilson}
\end{equation}
%Eq (19) Sudakov form factor as expectation value of Wilson line
where {\rm P} denotes path ordering along the integration contour, and
where ${<...>}_{A}$ denotes functional averaging in the gauge field
sector with whatever this entails (ghosts, gauge choice prescription,
Dirac determinant, etc.).
Having isolated a subsector of the full theory, where only gluons with
virtualities between $C_{1}/b$ and $C_{2}\xi Q$ are active degrees of
freedom, quark propagation and gluon emission can be described by
eikonal techniques, using either Feynman diagrams \cite{CS81,BS89} or
by employing a worldline casting of QCD which reverts the fermion
functional integral into a path integral \cite{GKKS97}.

Then the Sudakov functions, entering Eq. (\ref{eq:piamplbspace}), can be
expressed in terms of the momentum-dependent cusp anomalous dimension
of the bent contour to read
\begin{equation}
  s\left(\xi , b, Q, C_{1}, C_{2} \right)
=
  \frac{1}{2}
  \int_{C_{1}/b}^{C_{2}\xi Q}
  \frac{{\d}\mu}{\mu} \,
  \Gamma _{{\rm cusp}}
        \left(\gamma , \alpha _{\rm s}^{\rm an}(\mu )
        \right)
\label{eq:sudfuncusp}
\end{equation}
%Eq (20) Sudakov function in terms of cusp anomalous dimension
with the anomalous dimension of the cusp given by
\begin{equation}
  \Gamma _{{\rm cusp}}
        \left( \gamma , \alpha _{\rm s}^{\rm an}(\mu ) \right)
=
  2 \ln \left(
              \frac{C_{2}\xi Q}{\mu}
        \right) A\left( \alpha _{\rm s}^{\rm an}(\mu ) \right)
              + B\left( \alpha _{\rm s}^{\rm an}(\mu ) \right) \; ,
\label{eq:gammacusp}
\end{equation}
%Eq (21) Sudakov function in NLO
$\gamma = \ln \left( \frac{C_{2}\xi Q}{\mu} \right)$
being the cusp angle, i.e., the emission angle of a soft gluon and the
bent quark line after the external (large) momentum $Q$ has been
injected at the cusp point by the off-mass-shell photon.
The functions $A$ and $B$ are known at two-loop order:
\begin{eqnarray}
\everymath{\displaystyle}
  A\left( \alpha _{\rm s}^{\rm an}(\mu ) \right)
& = &
  \frac{1}{2}
  \left[
         \gamma _{\cal K}
                 \left( \alpha _{\rm s}^{\rm an}(\mu ) \right)
       + \beta (g) \frac{\partial}{\partial g}
       {\cal K}(C_{1}, \alpha _{\rm s}^{\rm an}(\mu ))
  \right]
\nonumber \\
& = &
       C_{\rm F} \frac{\alpha _{\rm s}^{\rm an}(g(\mu ))}{\pi}
     + \frac{1}{2} K\left( C_{1} \right) C_{\rm F}
       \left( \frac{\alpha _{\rm s}^{\rm an}(g(\mu ))}{\pi} \right)^{2}
\; ,
\label{eq:funAnlo}
\end{eqnarray}
%Eq (22) Function A in NLO
and
\begin{eqnarray}
\everymath{\displaystyle}
  B\left( \alpha _{\rm s}^{\rm an}(\mu ) \right)
& = &
  - \frac{1}{2}
    \left[
          {\cal K} \left(
                         C_{1}, \alpha _{\rm s}^{\rm an}(\mu )
                   \right)
        + {\cal G} \left(
                         \xi, C_{2}, \alpha _{\rm s}^{\rm an}(\mu )
                   \right)
    \right]
\nonumber \\
& = &
      \frac{2}{3} \frac{\alpha _{\rm s}^{\rm an}(g(\mu ))}{\pi}
      \ln \left(
          \frac{C_{1}^{2}}{C_{2}^{2}}\frac{{\rm e}^{2\gamma_{E}-1}}{4}
          \right) \; ,
\label{eq:funBnlo}
\end{eqnarray}
%Eq (23) Function B in NLO
respectively.
The K-factor in the ${\overline{\rm MS}}$ scheme to two-loop order
is given by \cite{Col89,KR87,CS81,DS84,KT82}
\begin{equation}
  K\left(C_{1}\right)
=
     \left(\frac{67}{18} - \frac{\pi ^{2}}{6}\right) C_{\rm A}
   - \frac{10}{9}n_{\rm f} T_{\rm F}
   + \beta _{0} \ln \left( C_{1} {\rm e}^{\gamma _{\rm E}}/2 \right)
\label{eq:Kfactor}
\end{equation}
%Eq (24) K-factor in \bar{MS} scheme
with $C_{\rm A}=N_{\rm C}=3$, $n_{\rm f}=3$, $T_{\rm F}=1/2$,
$\gamma _{\rm E}$ being the Euler-Mascheroni constant.

The quantities ${\cal K}$, ${\cal G}$ are calculable using the
nonabelian extension to QCD \cite{CS81} of the Grammer-Yennie method
\cite{GY73} for QED. Alternatively, one can calculate the cusp anomalous
dimension employing Wilson lines \cite{KR87,Kor89,GKKS97,KR92}.
In this latter approach (see, e.g., \cite{GKKS97}), the IR behavior of
the cusped Wilson line is expressed in terms of an effective fermion
vertex function whose variance with the momentum scale is governed by
the anomalous dimension of the cusp within the isolated effective
subsector. Since this scale dependence is strictly restricted within
the low-energy sector of the full theory, IR scales are locally coupled
and the soft form factor depends only on the cusp angle which varies
with the interquark transverse distance $b$ between $C_{1}/b$ and
$C_{2}\xi Q$.

The corresponding anomalous dimensions, are linked to each
other (for a nice discussion, see, \cite{Col89}) through the relation
$
  2 \Gamma _{{\rm cusp}} \left( \alpha _{\rm s}^{\rm an}(\mu ) \right)
=
 \gamma _{\cal K}\left( \alpha _{\rm s}^{\rm an}(\mu ) \right)
$
with
$
 \Gamma _{{\rm cusp}} (\alpha _{\rm s}^{\rm an}(\mu ))
=
  C_{\rm F}\, \alpha _{\rm s}^{\rm an}(\mu ^{2})/\pi
$, which shows that
$
 \frac{1}{2}\gamma _{\cal K}
=
 A\left( \alpha _{\rm s}^{\rm an}(\mu ) \right)
$.
(Note that $\gamma _{\cal G} = - \gamma _{\cal K}$.)

The soft amplitude
${\cal P}_{\pi}\left( x, b, C_{1}/b, \mu \right)$
and the hard-scattering amplitude
$T_{\rm H}\left( x, y, b, Q, \mu \right)$
satisfy independent RG equations to account for the dynamical
factorization (recall that both $b$ and $\xi$ are integration variables)
with solutions controlled by the modified ``evolution time'' (see,
e.g., \cite{DDT80} and earlier references cited therein):
\begin{eqnarray}
\everymath{\displaystyle}
  \tau \left( \frac{C_{1}}{b}, \mu \right)
& = &
  \int_{C_{1}^{2}/b^{2}}^{\mu ^{2}} \frac{{\d}k^{2}}{k^{2}} \,
  \frac{\alpha _{\rm s}^{{\rm an}(1)}(k^{2})}{4\pi}
\nonumber \\
& = &
  \frac{1}{\beta _{0}}
  \ln \frac{\ln\left(\mu ^{2}/\Lambda ^{2}\right)}
           {\ln\left(C_{1}^{2}/\left( b\Lambda \right)^{2}
               \right)}
+ \frac{1}{\beta _{0}}
  \left[
  \ln \frac{\mu ^{2}}{\left( C_{1}/b \right)^{2}}
- \ln \frac{\left\vert\mu ^{2} - \Lambda ^{2}\right\vert}
           {\left\vert\frac{C_{1}^{2}}{b^{2}} - \Lambda ^{2}\right\vert}
  \right]
\label{eq:evoltime}
\end{eqnarray}
%Eq (25) Modified evolution time
from the factorization scale
$C_{1}/b$ to the observation scale $\mu$, with $\Lambda$
denoting $\Lambda _{\rm QCD}$ as before. The evolution time is
directly related to the quark anomalous dimension, viz.,
$
 \gamma _{\rm q}\left( \alpha _{\rm s}^{\rm an}(\mu ) \right)
=
 - \alpha _{\rm s}^{{\rm an}}(\mu ^{2})/\pi
$.
One appreciates that the second term in (\ref{eq:evoltime}) stems
from the power-generated correction of the effective coupling and is
absent in the conventional approach. At moderate values of $\mu ^{2}$
it is ``slowing down'' the rate of evolution.

The leading contribution to the IR-modified Sudakov functions
$s\left( \xi ,b, Q, C_{1}, C_{2} \right)$
(where $\xi = x, \bar{x}, y, \bar{y}$)
is obtained by expanding the functions $A$ and $B$ in a power
series in $\alpha _{\rm s}^{\rm an}$, and collecting together all
large logarithms
$
 \left( \frac{\alpha _{\rm s}^{\rm an}}{\pi} \right)^{n}
 \ln \left ( \frac{C_{2}}{C_{1}} \xi b Q \right)^{m}
$,
which can be transformed back into large logarithms
$
 \ln \left( Q^{2}/{\kv}^{2} \right)
$
in transverse momentum space.
Employing equations (\ref{eq:oneloopalpha_an}) and
(\ref{eq:twoloopalpha_an}), the leading contribution results from the
expression
\begin{eqnarray}
\everymath{\displaystyle}
  s\left( \xi , b, Q, C_{1}, C_{2}\right)
= &&
    \frac{1}{2}
    \int_{C_{1}/b}^{C_{2}\xi Q} \frac{{\d}\mu}{\mu}
    \Biggl\{
           2 \ln \left( \frac{C_{2} \xi Q}{\mu} \right)
       \Biggl[
         \frac{\alpha _{\rm s}^{{\rm an}(2)}(\mu )}{\pi}
         A^{(1)}
      +  \left(
         \frac{\alpha _{\rm s}^{{\rm an}(1)}(\mu )}{\pi}
         \right)^{2}
         A^{(2)}\left( C_{1} \right)
       \Biggr]
\nonumber \\
&&    +  \frac{\alpha _{\rm s}^{{\rm an}(1)}(\mu )}{\pi}
         B^{(1)}\left( C_{1}, C_{2} \right)
      + {\cal O} \left( \frac{\alpha _{\rm s}^{\rm an}}{\pi}\right)^{3}
    \Biggr\} \; ,
\label{eq:SudakovNLO}
\end{eqnarray}
%Eq (26) Sudakov at NLO in IR-finite scheme
where Eq. (\ref{eq:twoloopalpha_an}) is to be used in front of
$A^{(1)}$, whereas the other two terms are to be evaluated with
Eq. (\ref{eq:oneloopalpha_an}). The specific values of the coefficients
$A^{(i)}$, $B^{(i)}$ are
\begin{eqnarray}
\everymath{displaystyle}
  A^{(1)}                         & = & C_{\rm F}
\nonumber \\
  A^{(2)}\left(C_{1}\right)       & = & \frac{1}{2}\, C_{\rm F}\,
                                        K\left( C_{1} \right)
\nonumber \\
  B^{(1)}\left(C_{1},C_{2}\right) & = & \frac{2}{3}
                \ln \left(
                          \frac{C_{1}^{2}}{C_{2}^{2}}
                          \frac{{\rm e}^{2\gamma_{E}-1}}{4}
                    \right) \; .
\label{eq:expcoef}
\end{eqnarray}
%Eq (27) Expansion coefficients for the Sudakov function up to NLLO
As now the power-correction term in $\alpha _{\rm s}$ gives rise to
polylogarithms, a formal analytic expression for the Sudakov form factor
is too complicated for being presented. In the calculations to follow,
Eq. (\ref{eq:SudakovNLO}) is evaluated numerically with appropriate
kinematic bounds. Note that, neglecting the power-generated logarithms,
we obtain an equation for the ordinary Sudakov function, which we write
as an expansion in inverse powers of $\beta _{0}$ to read
\begin{eqnarray}
\everymath{\displaystyle}
  s\left( \xi , b, Q, C_{1}, C_{2} \right)
= &&
  \frac{1}{\beta _{0}}
  \left[
         \left(
               2 A^{(1)} \hat Q + B^{(1)}
         \right)\ln \frac{\hat Q}{\hat b}
         -2 A^{(1)} \left( \hat Q - \hat b \right)
  \right]
\nonumber \\
&& - \, \frac{4}{\beta _{0}^{2}}
     A^{(2)}
     \left(
           \ln \frac{\hat Q}{\hat b} -
           \frac{\hat{Q} - \hat{b}}{\hat b}
     \right)
\nonumber \\
&& + \frac{\beta _{1}}{\beta _{0}^{3}} A^{(1)}
     \left\{
              \ln \frac{\hat Q}{\hat b}
            - \frac{\hat Q - \hat b}{\hat b}
              \left[
                    1 + \ln \left( 2 \hat b \right)
              \right]
            + \frac{1}{2}
              \left[
                     \ln ^{2}\left(2\hat Q \right)
                   - \ln ^{2}\left(2\hat b \right)
              \right]
     \right\} \; ,
\label{eq:oldsudakov}
\end{eqnarray}
%Eq (28) Corrected expression for the ordinary Sudakov function
where the abbreviations \cite{LS92}
$
 \hat Q
\equiv
 \ln \frac{C_{2} \xi Q}{\Lambda}
$
and
$
 \hat b
\equiv
  \ln \frac{C_{1}}{b\Lambda}
$
have been used.

This quantity differs from the original result given by Li and Sterman
in \cite{LS92}, and, though it almost coincides numerically with the
formula derived by J. Bolz \cite{Bol95}, it differs from that
algebraically.

All told, the final expression for the electromagnetic pion form factor
at leading perturbative order in $T_{\rm H}$ and next-to-leading
logarithmic order in the Sudakov form factor has the form
\begin{eqnarray}
\everymath{\displaystyle}
  F_{\pi}(Q^{2})
= &&
  \frac{2}{3}\, \pi\, C_{\rm F}\, f_{\pi}^{2}\,
  \int_{0}^{1} {\d}x
  \int_{0}^{1} {\d}y
  \int_{0}^{\infty}b\, {\d}b\,
  \alpha _{\rm s}^{{\rm an}(1)}\left(\mu _{\rm R}\right)
  \phi _{\rm as} (x)
  \phi _{\rm as} (y)
  \exp \left[-\frac{b^{2} \left(x\bar{x} + y\bar{y}\right)}
  {4\beta _{\rm as}^{2}}\right]
\nonumber \\
&& \times
  K_{0}\left(\sqrt{xy}Qb\right)
   \exp \left[ - S \left(x,y,b,Q, C_{1},C_{2},C_{4}\right) \right] \; ,
\label{eq:pifofafin}
\end{eqnarray}
%Eq (29) Final form for electromagnetic form factor of pion
where
\begin{equation}
  S \left(x,y,b,Q, C_{1},C_{2},C_{4}\right)
\equiv
   s\left(x, b, Q, C_{1}, C_{2}\right)
 + s\left(\bar{x}, b, Q, C_{1}, C_{2}\right)
 + (x \leftrightarrow y)
 - 8 \,\tau \left(\frac{C_{1}}{b}, \mu _{\rm R}\right)
\label{eq:fullSudakov}
\end{equation}
%Eq (30) Full Sudakov expression, i.e., including evolution
with $\tau \left(\frac{C_{1}}{b}, \mu _{\rm R}\right)$ given by
Eq. (\ref{eq:evoltime}) and $\mu _{\rm R} = C_{4}f(x,y)Q$.

Before we go beyond the leading order in the perturbative expansion
of the hard-scattering amplitude, $T_{\rm H}$, let us pause for a moment
to comment on the pion wave function. We have proactively indicated in
Eq. (\ref{eq:pifofafin}) that its asymptotic form
$\phi _{\rm as}(x) = 6 x \bar x$
will be used.

A few words about this choice are now in order.

Hadron wave functions are clearly the essential variables needed to
model and describe the properties of an intact hadron.
In the past, most attempts to improve the theoretical predictions for
the hard contribution to the pion form factor have consisted of using
end-point concentrated wave functions. In this analysis, we refrain
from using such wave functions of the Chernyak-Zhitnitsky (CZ) type
\cite{CZ84}, referring for a compilation of objections and references to
\cite{SSK98} (see also \cite{Rad98}), and present instead evidence for
an alternative source of enhancement due to the nonperturbative power
correction in the effective coupling.
This effect is found \cite{SSK98} to be quite significant, even for the
asymptotic pion wave function which has its maximum at $x=1/2$.
Indeed, the IR-enhanced hard contribution can account already at
leading perturbative order for a sizable part of the measured magnitude
of the electromagnetic pion form factor, though agreement with the data
calls for the inclusion of the soft contribution
(cf. Eq. (\ref{eq:softhard})) \cite{IL84,Rad90,BH94,JKR96} -- even if
the NLO correction is taken into account (see sect. \ref{sec:valid}).
Nevertheless, the true pion distribution amplitude may well be a
``hybrid'' of the type
\hbox{
$
    \Phi _{\pi}^{\rm true}
 =
    90\% \Phi _{\pi}^{\rm as}
  + 9\% \Phi _{\pi}^{\rm CZ}
  + 1\% C_{4}^{(3/2)}
$},
where the mixing ensures a broader shape with the fourth-order,
``Mexican hat''-like, Gegenbauer polynomial $C_{4}^{(3/2)}$, being added
in order to cancel the dip of $\phi _{\rm CZ}$ at $x=1/2$.

First tasks from instanton-based approaches show that the extracted
pion distribution amplitudes are very close to the asymptotic form
\cite{Dor96,PPRWG98}.
Similar results are also obtained using nonlocal condensates
\cite{BM95}.
The discussion of non-asymptotic pion distribution amplitudes will be
conducted in a separate publication.

\section{P\lowercase{ion form factor to order}
$\left(\alpha _{\lowercase{\rm s}}^{\lowercase{\rm an}}
               (Q^{2})\right)^{2}$}
\label{sec:piffNLO}

Next, we generalize our calculation of the hard contribution to the pion
form factor by taking into account the perturbative correction to
$T_{\rm H}$ of order $\alpha _{\rm s}^{2}$, using the results obtained
in \cite{FGOC81,DR81,BT87,MNP98}, in combination with our analytical
scheme.

To be precise, we only include the NLO corrections to $T_{\rm H}$
leaving NLO corrections to the evolution of the pion distribution
amplitude aside. The reason is that for the asymptotic distribution
amplitude, at issue here, these corrections are tiny
\cite{Mue94,MNP98}.
For subasymptotic distribution amplitudes, however, NLO evolutional
corrections \cite{Mue94} have to be taken into account.
The calculation below does not claim to be strictly correct, the
reason being that the transverse degrees of freedom in the NLO terms to
$T_{\rm H}$ have been neglected, albeit the intrinsic ones in the
wave functions are taken into account. Hence our prediction should be
regarded rather as an {\it upper limit} for the size of the hard
contribution to the pion form factor than as an exact result. Taking
into account the $k_{\perp}$-dependence of $T_{\rm H}$ at NLO, this
result will be reduced, though we expect that due to IR enhancement this
reduction should be rather small. Thus, we have
\begin{eqnarray}
\everymath{displaystyle}
  F_{\pi}\left( Q^{2} \right)
& = &
  16\pi C_{\rm F}
  \left(
        \frac{f_{\pi}}{2\sqrt{N_{\rm C}}!}
  \right)^{2}
  \int_{0}^{1} {\d}x \int_{0}^{1} {\d}y
  \int_{0}^{\infty} b \,{\d}b \,
  \alpha_{\rm s}^{\rm an}\left( \mu_{\rm R}^{2}\right)
\nonumber \\
&& \times
  K\left(\sqrt{x y}Qb\right)
  \phi _{\rm as}(x) \phi _{\rm as}(y)
\nonumber \\
&& \times
  \exp\left(
            - \frac{b^{2}\left( x\bar x + y\bar y \right)}
              {4\beta _{\rm as}^{2}}
      \right)
  \exp\left( - S\left(x, y, b, Q, C_{1}, C_{2}, C_{4} \right)
      \right)
\nonumber \\
&& \times
  \left[
        1 + \frac{\alpha _{\rm s}^{\rm an}}{\pi}
            \left( f_{\rm UV}\left( x, y, Q^{2}/\mu _{\rm R}^{2} \right)
          + f_{\rm IR}\left(x, y, Q^{2}/\mu _{\rm F}^{2}\right)
          + f_{\rm C}(x, y)
            \right)
  \right] \; ,
\label{eq:NLOpiff}
\end{eqnarray}
%Eq (31) Pion form factor in NLO for T_H
where the Sudakov form factor, including evolution, is given by
Eq. (\ref{eq:fullSudakov}), $\mu _{\rm F}= C_{1}/b$, and the functions
$f_{i}$ are taken from \cite{MNP98}
\footnote{We wish to thank Dr.~B.~Meli\v c for sending us the corrected
version of this paper prior to publication.}.
They are given by
\begin{eqnarray}
\nonumber
  f_{\rm UV}\left(x, y, Q^{2}/\mu _{\rm R}^{2}\right)
& = &
  \frac{\beta _{0}}{4}
    \left(\frac{5}{3}
  - \ln\left(\bar{x}\bar{y}\right)
  + \ln\frac{\mu _{\rm R}^{2}}{Q^{2}}\right) \\
\nonumber
  f_{\rm IR}\left(x, y, Q^{2}/\mu _{\rm F}^{2}\right)
& = &
  \frac{2}{3}
  \left(
        3 + \ln\left(\bar{x}\bar{y}\right)
  \right)
         \left(
                 \frac{1}{2}\ln\left(\bar{x}\bar{y}\right)
               - \ln\frac{\mu _{\rm F}^{2}}{Q^{2}}
         \right) \\
\nonumber
  f_{\rm C}(x,y)
& = &
  \frac{1}{12}
  \left[- 34 + 12 \ln\left(\bar{x}\bar{y}\right)
        + \ln x \ln y \right. \\
\nonumber
&& + \ln\bar{x}\ln\bar{y} - \ln x\ln\bar{y} - \ln\bar{x}\ln y \\
&& \left. + (1-x-y)H(x,y) + R(x,y)\right]
\label{eq:fis}
\end{eqnarray}
%Eq (32) Definition of the functions f_i entering the NLO calculation
and are related to UV and IR poles, as indicated by corresponding
subscripts, that have been removed by dimensional regularization along
with the associated constants
$
 \ln (4\pi ) - \gamma _{\rm E}
$, whereas the last term is scale-independent.
In evaluating expression $f_{\rm C}$ in (\ref{eq:fis}), we found it
particularly convenient to use the representation of the function
$H(x,y)$ given by Braaten and Tse \cite{BT87},
\begin{equation}
  H(x,y)
=
  \frac{1}{1-x-y}
  \left[
          {\rm Li}_{2}(\bar x)
        + {\rm Li}_{2}(\bar y)
        - {\rm Li}_{2}(x)
        - {\rm Li}_{2}(y)
        + \ln x \ln y
        - \ln \bar x \ln \bar y
  \right] \; ,
\label{eq:H}
\end{equation}
%Eq (33) Braaten and Tse's expression for function H(x,y)
where ${\rm Li}_{2}$ denotes the Spence function.
For the function $R(x,y)$ we have used the expression derived by Field
et al. \cite{FGOC81}, except at point $x\approx y$, where we employed
the Taylor expansion displayed below:
\begin{eqnarray}
\everymath{dispalystyle}
  R(x,y)
& = &
  \frac{1}{3\,\left( -1 + y \right)\, y^{2}}
  \Bigl[
        \left( - 1 + 33\,y - 45\, y^{2} + 13\,y^{3} \right) \,
          \ln \bar y
\nonumber \\
&& \phantom{\frac{1}{3\,\left(-1+y\right)\,y^{2}}}
               + y\,\left( -1 + y + ( 9 - 13\,y )\,y\,\ln y
        \right)
   \Bigr]
\nonumber \\
&&
 + \frac{x-y}{ 3\,(-1+y)^{2}\,y^{3} }
   \Bigl[
         ( - 1 + y )^{2}\, ( -1 + 16\,y )\, \ln \bar y
\nonumber \\
&& \phantom{\frac{x-y}{ 3\,(-1+y)^{2}\,y^{3}}+}
           + y\, \left( -1 + 13\,y - 12\,y^{2} + 2\,y^{2}\,\ln y
                 \right)
   \Bigl]
\nonumber \\
&&
 + \frac{ ( x - y )^{2} }{ 30\, ( -1 + y )^{3}\,y^{4} }
    \Bigl[
           ( - 1 + y )^{3}\, \left( 9 - 148\,y + 9\,y^{2}\right)\,
           \ln \bar y
\nonumber \\
&& \phantom{\frac{(x-y)^{2}}{30\,(-1+y)^{3}\,y^{4}}+}
          - y \left( 9 - 148\,y + 328\,y^{2} - 189\,y^{3}
          + y^{3}\,( 5 + 9\,y )\,\ln y \right)
    \Bigr] \; .
\label{eq:TaylorR}
\end{eqnarray}
%Eq (34) Taylor expansion around x=y of R(x,y)
Note that this expression does not reproduce its counterpart in
\cite{FGOC81}.

Having developed in detail the theoretical apparatus, let us now turn to
the concrete calculation of the pion form factor up to NLO.

\section{V\lowercase{alidity of the analysis}}
\label{sec:valid}

This numerical analysis updates and generalizes our previous
investigation in \cite{SSK98}.
In order to set up a reliable algorithm for the numerical evaluation of
the expressions presented above, we have to ensure that this is done in
kinematic regions where use of fixed-order or resummed perturbation
theory is legal.
Further, restrictions have to be imposed to avoid double counting of
gluon corrections by carefully defining the validity domain of each
contribution to the pion form factor.
These kinematic constraints are compiled below.

\begin{center}
{\small{\it  Kinematic cuts}}
\end{center}
\begin{enumerate}
\item $C_{1}/b > \Lambda _{\rm QCD}$; otherwise the whole Sudakov
      exponent $\exp (-S)$ (cf. Eq. (\ref{eq:fullSudakov})) is continued
      to zero because this large-$b$ region is properly taken into
      account in the wave functions. This condition excludes from
      the resummed perturbation theory soft gluons with wavelengths
      larger than $C_{1}/\Lambda$, which should be treated
      nonperturbatively.
\item $C_{2}\xi Q > C_{1}/b$; otherwise each Sudakov exponent
      $\exp \left[- s\left(\xi,b,Q,C_{1},C_{2}\right)\right]$ in
      Eq. (\ref{eq:fullSudakov}) is ``frozen'' to unity because this
      small-$b$ region is dominated by low orders of perturbation theory
      rather than by the resummed perturbation series, and consequently
      contributions in this region should be ascribed to higher-order
      corrections to $T_{\rm H}$.
      Yet evolution is taken into account to match the scales in our
      ``gliding'' factorization scheme.
\item $C_{4}f(x,y)Q > C_{1}/b$; otherwise evolution time
      $\tau \left( C_{1}/b, \mu _{\rm R} \right)$ in
      Eq. (\ref{eq:fullSudakov}) is contracted to zero, i.e., evolution
      is ``frozen''. The renormalization scale should be at least equal
      to the factorization scale, so that the effective coupling has
      arguments in the range controlled by perturbation theory.
\item $C_{4}f(x,y)Q > C_{2}\xi Q$; otherwise evolution to that scale
      is ``frozen'' because this region is appropriately accounted for
      by the Sudakov contribution. This helps avoiding double counting
      terms which belong to the resummed rather than to the fixed-order
      perturbation theory.
\item $C_{4}f(x,y)Q > C_{1}/b$; otherwise the two scales
      $\mu _{\rm R}=\mu _{\rm F}=C_{1}/b$
      are identified in the function $f_{\rm UV}(x,y)$.
      If $\mu _{\rm R}\leq \Lambda _{\rm QCD}$, then
      $f_{\rm UV}(x,y)$ is set equal to zero.
\item $C_{1}/b > \Lambda _{\rm QCD}$; otherwise the function
      $f_{\rm IR}(x,y)$ is set equal to zero.
\end{enumerate}

\begin{figure}
\tighten
\[
\psfig{figure=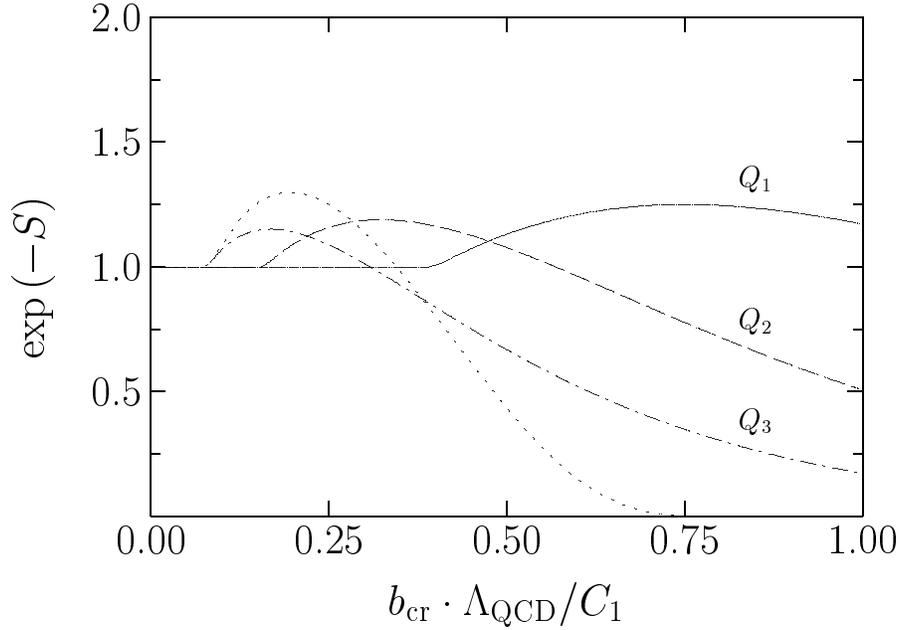,%
      bbllx=148pt,bblly=480pt,bburx=390pt,bbury=690pt,%
      height=7.0cm,silent=}
\]
\vspace{1.5cm}
\caption[fig:sudall]
        {\tenrm Behavior of the Sudakov form factor with respect to
         the transverse separation b for three different values of
         the momentum transfer
         $Q_{1}=2$~GeV, $Q_{2}=5$GeV, and $Q_{3}=10$~GeV with all
         $\xi _{i}=1/2$, and where we set
         $C_{1}=2{\rm e}^{-\gamma _{\rm E}}$,
         $C_{2}={\rm e}^{-1/2}$ and
         $\Lambda _{\rm QCD}=0.242~{\rm GeV}$. The dotted curve shows
         the result obtained with $\alpha _{s}^{\overline{MS}}$ and
         $\Lambda _{\rm QCD}=0.2~{\rm GeV}$ for $Q_{2}=5$~GeV using the
         same set of $C_{i}$ as before. Notice that in this case,
         evolution is limited by the (renormalization) scale
         $\mu_{\rm R}=t=\{ {\rm max}\sqrt{xy}\, Q, C_{1}/b \}$,
         proposed in \cite{LS92}.
         However, the enhancement at small $b$-values due to the quark
         anomalous dimension is not neglected.
\label{fig:sudall7}}
\end{figure}

To illustrate the difference in technology between approaches
employing the conventional expression for the full Sudakov exponent
\cite{BS89,LS92}, on one hand, and our analysis, on the other hand, we
show $\exp (-S)$ graphically in Fig. \ref{fig:sudall7} for three
different values of the momentum transfer. In contrast to Li and Sterman
\cite{LS92}, the evolutional contribution is not cut-off at unity,
whenever $C_{2}\xi Q < C_{1}/b$.
The dotted curve shows the result for Eq. (\ref{eq:oldsudakov}) without
this cutoff. One infers from this figure that their suggestion to ignore
the enhancement due to the anomalous dimension does not apply in our
case because the IR-modified Sudakov form factor is not so rapidly
decreasing as $b$ increases owing to the IR-finiteness of
$\alpha _{\rm s}^{\rm an}$.
Indeed, as $Q$ becomes smaller, $\exp (-S)$ remains constant and fixed
to unity for increasing $b$, providing enhancement only in the
large-$b$ region before it reaches the kinematic boundary
$C_{1}/b=\Lambda _{\rm QCD}$, where it is set to zero.
As a result, for small $Q$-values, like $Q_{1}=2$~GeV, the enhancement
due to the quark anomalous dimension cannot be associated with
higher-order corrections to $T_{\rm H}$, since it operates at larger
$b$-values, and for that reason it should be taken into account.
Only for asymptotically large $Q$ values, when the IR-modified Sudakov
form factor and the conventional one become indistinguishable, the
evolutional enhancement becomes a small effect, strictly confined in the
small-$b$ region, and can be safely ignored.

%-----------------------------------------------------------------------
%                            F I G U R E  2
%                         \label{fig:sudall7}
%-----------------------------------------------------------------------
%
%---------------------------------------------------------------------

This behavior of the Sudakov form factor, we emphasize, shows that
power-induced subleading logarithmic corrections are relevant in the
range of currently probed momentum-transfer values. Then the advantage
of employing such a scheme to calculate observables is that the hard
contribution to the pion form factor is enhanced.
This is because the range in which soft gluons build up the Sudakov form
factor is enlarged and inhibition of bremsstrahlung sets in at larger
$Q^{2}$. Let us mention in this context that power corrections could
also lead to additional suppression of soft gluon emission at large
transverse distance $b$. Indeed, Akhoury, Sincovics, and Sotiropoulos
\cite{ASS98} have resummed nonperturbative power corrections in $b$ to
the hadronic (meson) wave function and found Sudakov-type suppression
on top of the Sudakov suppression discussed so far. The discussion of
such IR-renormalon-based contributions in conjunction with our approach
would be interesting, but goes beyond the scope of the present work.

\begin{figure}
\tighten
\centerline{\epsfig{file=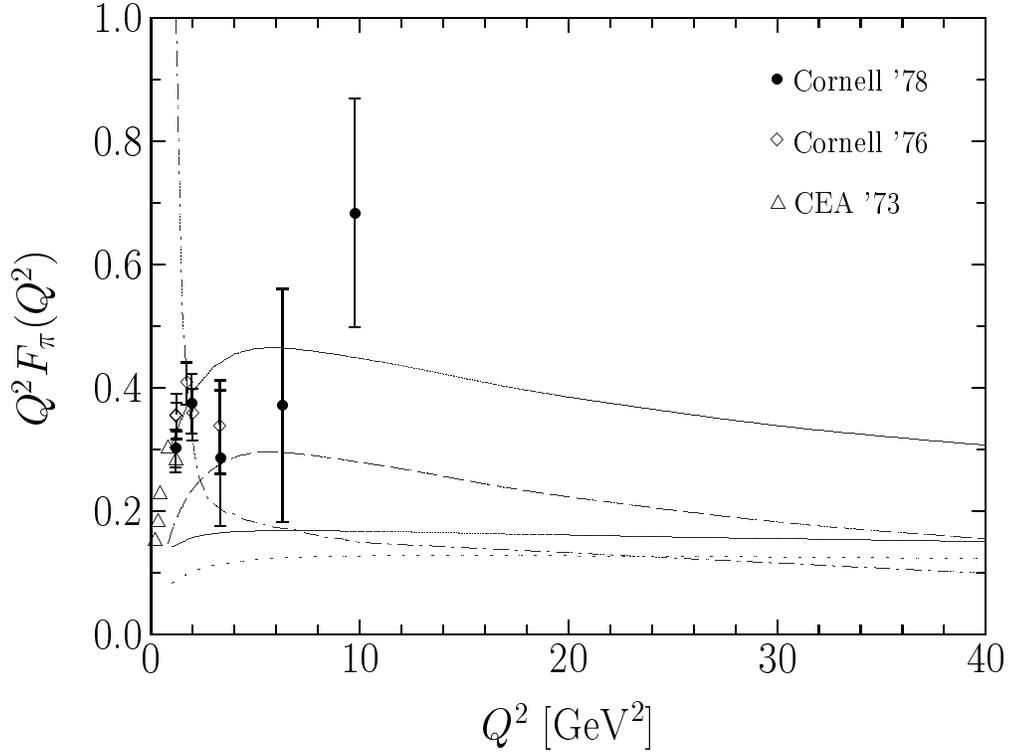,height=10cm,width=14cm,silent=}}
\vspace{1cm}
\caption[fig:pionffall]
        {\tenrm Spacelike pion form factor calculated with
         $\phi _{\rm as}$ within our theoretical scheme in comparison
         with existing experimental data \cite{Bro73,Beb76}. The dotted
         line shows the enhanced hard contribution to the form factor,
         obtained in our analysis at leading perturbative order and
         with inclusion of the NLO correction to $T_{\rm H}$ (lower
         solid line). The upper solid line represents the sum of the
         lower solid line and the dashed line, the latter curve giving
         the result for the soft contribution, computed in \cite{JKR96}.
         The dot-dashed line reproduces the recent calculation of Tung
         and Li \cite{TL98} which does not include an intrinsic
         $k_{\perp}$-dependence and uses for $\alpha _{\rm s}$ the
         conventional 1-loop expression.
\label{fig:piff06}}
\end{figure}

Using the techniques discussed above, we obtain the theoretical
predictions shown in Fig. \ref{fig:piff06}. Recall that the asymptotic
pion distribution amplitude $\phi _{\rm as}=6x(1-x)$ is employed with
$\beta _{\rm as}^{2}=0.883~[{\rm GeV}^{-2}]$ which corresponds to
$ \langle \kv ^{2} \rangle^{1/2}= 0.35$~GeV (see, \cite{JK93}).
A set of constants \hbox{$C_{i}$, $(i=1,2,3)$} which eliminates
artifacts of dimensional regularization, while practically preserving
the matching between the resummed and the fixed-order calculation, is
given in table \ref{tab:setsC_i} in comparison with other common choices
of these constants. In addition, we set $f(x,y)=\sqrt{xy}$, a choice
for the renormalization scale which eliminates large kinematical
corrections due to soft gluon emission.
However, these favored values of the scheme constants do not constitute
a strict constraint on the validity of the numerical analysis. They
merely indicate the anticipated appropriate choice of the factorization
and renormalization scales with respect to observables and theoretical
self-consistency.

\begin{table}
\caption[tab:coefc_i]
        {Different sets of coefficients $C_{i}$, and values of the
         $K$-factor and the quantity (cf. Eq.~(\ref{eq:funBnlo}))
         $\kappa = \ln \left(
          C_{1}^{2}\;{\rm e}^{2\gamma_{E}-1} /4\, C_{2}^{2}
                       \right)$,
         corresponding to different factorization and renormalization
         prescriptions.
\label{tab:setsC_i}}
\begin{tabular}{lcccccc}
            & \multicolumn{4}{c}{Scheme parameters $C_{i}$} \\
 Choice & $C_{1}$ &
 $C_{2}=\frac{1}{\sqrt{2}}C_{2}^{\rm CSS}$ &
 $C_{3}$ & $C_{4}$ & $K$ & $\kappa$ \\
\hline
 canonical & $2\exp \left(-\gamma _{\rm E}\right)$ & $1/\sqrt{2}$ &
 $2\exp \left(-\gamma _{\rm E}\right)$ & -- & 4.565 & -0.307 \\
 SSK \cite{SSK98} &
 $\exp\left[-\frac{1}{2}\left(2\gamma _{\rm E}-1\right)\right]$ &
 $1/\sqrt{2}$ &
 $\exp\left[-\frac{1}{2}\left(2\gamma _{\rm E}-1\right)\right]$ & -- &
 2.827 & 0 \\
 this work & $2\exp \left(-\gamma _{\rm E}\right)$ &
 $\exp \left(-1/2\right)$ & $2\exp \left(-\gamma _{\rm E}\right)$ &
 $\exp \left(-1/2\right)$ & 4.565 & 0 \\
\end{tabular}
\end{table}
%T A B L E  1

%-----------------------------------------------------------------------
%                            F I G U R E  3
%                          \label{fig:piff06}
%-----------------------------------------------------------------------
%
%---------------------------------------------------------------------

\begin{figure}
\tighten
\centerline{\epsfig{file=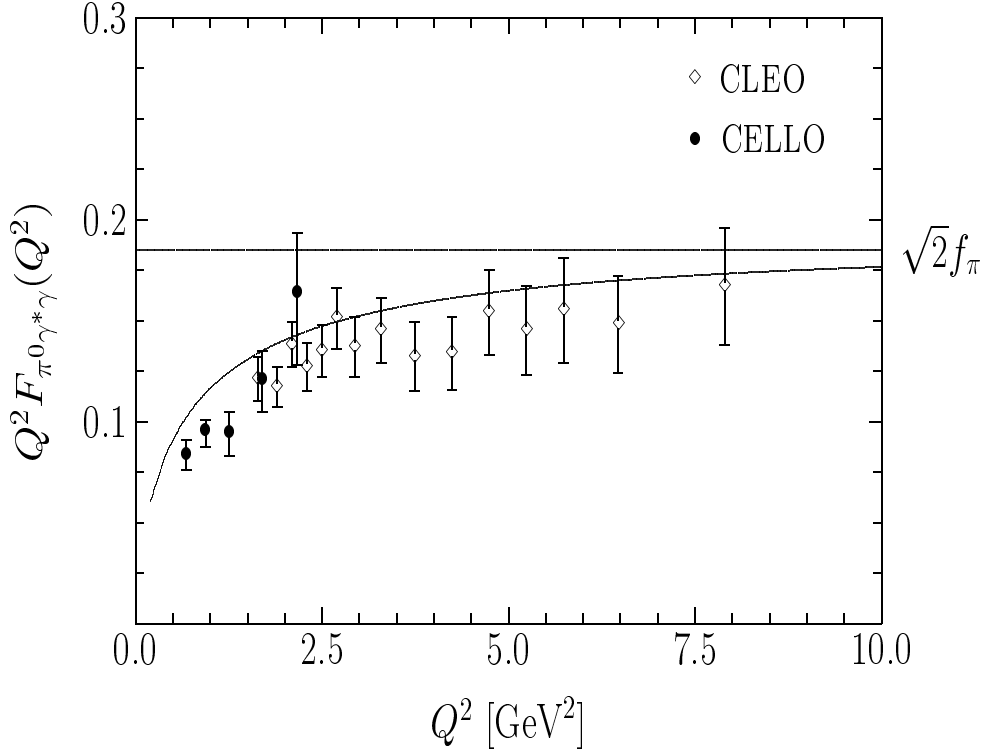,height=10cm,width=14cm,silent=}}
\vspace{1cm}
\caption[fig:piongamma]
        {\tenrm Pion-photon transition form factor calculated with
        $\phi _{\rm as}$ and IR enhancement (lower solid line).
        The horizontal line represents the asymptotic behavior.
        The data are taken from \cite{CLEO98,CELLO91}.
\label{fig:piga26}}
\end{figure}

Before we proceed with the discussion of these results, let us first
present the theoretical prediction for the pion-photon transition form
factor $F_{\pi\gamma ^{*}\gamma}(Q^{2},q^{2}=0)$ in which one of the
photons is highly off-shell and the other one is close to its
mass-shell.
In leading perturbative order this form factor is given by the
expression (cf. \cite{JKR96})
\begin{eqnarray}
\everymath{displaystyle}
  F_{\pi\gamma}\left(Q^{2}\right)
& = &
   \frac{1}{\sqrt{3}\pi}
   \int_{0}^{1} {\d}x \int_{0}^{\infty} {\d}b \,
   b \, \frac{f_{\pi} \phi _{\rm as}(x)}{2\sqrt{6}}
   \exp\left(-x\bar{x}b^{2}/4\beta _{\rm as}^{2}
       \right)
\nonumber \\
&& \times
   \left(4\pi K_{0}\left(\sqrt{\bar{x}}\,bQ\right)\right)
   e^{-S_{\pi\gamma}} \; ,
\label{eq:ffpigamma}
\end{eqnarray}
%Eq (35) Pion-photon transition form factor
where the Sudakov exponent, including evolution, has the form
\begin{equation}
  S_{\pi\gamma}\left( x, \bar{x}, b, Q, C_{1}, C_{2}, C_{4} \right)
=
    s(x,b,Q) + s(\bar{x},b,Q)
  - 4 \tau \left( \frac{C_{1}}{b}, \mu _{\rm R} \right) \; .
\label{eq:sudpigamma}
\end{equation}
%Eq (36) Full Sudakov exponent for pion-photon transition form factor
The main difference relative to the previous case is that this form
factor contains only one pion wave function, whereas the associated
hard-scattering part, being purely electromagnetic at this order, does
not depend on $\alpha _{s}$.
The only dependence on the strong coupling constant enters through the
anomalous dimensions in the Sudakov form factor. The result of this
calculation is displayed in Fig. \ref{fig:piga26}.

%-----------------------------------------------------------------------
%                            F I G U R E  4
%                         \label{fig:piga26}
%-----------------------------------------------------------------------
%
%---------------------------------------------------------------------

All constraints on kinematics set forward in the numerical evaluation
of the electromagnetic pion form factor are relevant to this case too,
except the requirement which deals specifically with the choice of the
renormalization scale, which now is set equal to
$\mu _{\rm R} = C_{4} x Q$. Another reasonable choice would be
$\mu _{\rm R} = C_{4} \sqrt{x\bar x} Q$, which entails evolution to a
lower scale, hence reducing evolutional enhancement through
$\tau \left(C_{1}/b,\mu _{\rm R} \right)$ by approximately $6\%$.

It is obvious from Fig. \ref{fig:piff06} that the IR-enhanced hard
contribution to $F_{\pi}(Q^{2})$ is providing a sizeable fraction of
the magnitude of the form factor. This behavior is IR stable and
extends from low to high $Q^{2}$ values, exhibiting exact scaling.
In contrast to the dot-dashed line, which is calculated with the
conventional $\alpha _{\rm s}^{(1)}$, enhancement is not resulting from
approaching the unphysical Landau singularity at
$Q^{2}=\Lambda _{\rm QCD}^{2}$ -- still appreciable at $Q^{2}$ values
around $(6-8)~{\rm GeV}^{2}$ -- but is induced by the nonperturbative
power correction.

To make these findings more transparent, we compare our results in
table \ref{tab:piffval} with those obtained in other analyses.

\begin{table}
\caption[tab:valpiff]
        {Calculated pion form factor at two values of $Q^{2}$. The
         first two columns show the results obtained in the present
         work, in comparison with those calculated by Jakob and Kroll
         (JK) \cite{JK93} (third column), and by Meli\v c et al. (MNP)
         \cite{MNP98} (last two columns).
\label{tab:piffval}}
\begin{tabular}{cccccc}
         $Q^{2}$~[GeV${}^{2}$] & LO & LO+NLO & LO & LO & LO+NLO\\
         & (this work) & (this work) & (JK) & (MNP) & (MNP) \\
\hline
         4  & 0.119 & 0.167 & 0.08 & 0.131 & 0.211 \\
        10  & 0.128 & 0.168 & 0.08 & 0.109 & 0.164 \\
\end{tabular}
\end{table}
%T A B L E  2

Comparison of our values with those calculated by Jakob and Kroll
\cite{JK93} shows that the suppression of the hard part of the form
factor due to the inclusion of transverse degrees of freedom is
counteracted by the power-induced enhancement. On the other hand,
comparison with the values computed by Meli\v c et al. at leading order,
by ignoring completely transverse degrees of freedom, reveals that in
the $Q^{2}$ domain, where the influence of the Landau singularity has
died out, there is enhancement. Furthermore, comparing our results with
theirs at next-to-leading order, we conclude that our choice of the
renormalization scale is consistent with a proper matching between
gluon corrections, calculated on a term-by-term perturbation expansion,
and those due to the resummed perturbative series. Moreover, the
scaling behavior of the calculated perturbative contribution to the full
pion form factor is also improved.

The calculation of $F_{\pi\gamma}(Q^{2})$ provides an additional
confirmation of our method. Fig. \ref{fig:piga26} shows that our
theoretical prediction for this form factor reproduces the recent
high-precision CLEO \cite{CLEO98} and the earlier CELLO \cite{CELLO91}
data with the same numerical accuracy as the dipole interpolation
formula, without using any phenomenological adjustment.

In this context, we mention, however, that other authors
\cite{RR96,MR97,Kho97} obtain similarly good numerical agreement of
$F_{\pi\gamma ^{*}}$ with the experimental data, following different
premises on the basis of QCD sum rules.

Another important point worth to be mentioned is that the proposed
scheme exceeds the quality of the phenomenological predictions derived
from schemes which involve saturation prescriptions \cite{JA90}, even
if these employ ``comensurate rescaling'' to ``pre-sum'' into the
running coupling constant NLO contributions \cite{BJPR98}. A more
detailed study of the BLM procedure \cite{BLM83}, which helps avoid
scheme ambiguities, in combination with our approach will be presented
elsewhere.

\section{S\lowercase{ummary and conclusions}}
\label{sec:sum}

We have developed in detail a theoretical framework which
self-consistently incorporates effects resulting from a modification
of the running $\alpha _{s}$ by a nonperturbative power correction
\cite{SS97} which provides IR universality.
Though a deep physical understanding of such contributions is still
lacking, we have given, as a matter of practice, quantitative evidence
that in this way it is possible to get a hard contribution to the
electromagnetic form factor $F_{\pi}(Q^{2})$, which is IR-enhanced
relative to conventional approaches, using solely the asymptotic form of
the pion distribution amplitude, hence avoiding end-point concentrated
distribution amplitudes. The presented IR-finite factorization
and renormalization scheme makes it possible to take into account
transverse degrees of freedom both in the pion wave function \cite{JK93}
as well as in the form of Sudakov effects \cite{LS92}, without entailing
suppression of the form-factor magnitude.
In addition, use of this modified form of $\alpha _{\rm s}(Q^{2})$
renders the theoretical predictions insensitive to its variation with
$Q^{2}$ at small momentum values. An appropriate choice of the
factorization and renormalization scales ensures matching of
fixed-order with resummed perturbation theory and helps avoid double
counting of higher-order corrections due to gluon radiation, ensuring
this way the self-consistency of the whole perturbative treatment.
The same procedure applied to $F_{\pi ^{0}\gamma ^{*}\gamma}$ yields a
prediction which agrees with the experimental data very well, hence
supporting the theoretical basis of our approach.

Taking our results for the pion form factor at face value, they seem
to suggest that its magnitude may be significantly larger than predicted
in previous analyses, even without employing broad pion distribution
amplitudes. However, because of the restricted accuracy of the
existing experimental data such a conclusion would be surely premature.

We believe that the insight gained through our analysis gives a strong
argument that if power corrections to the effective coupling of QCD
really exist, their implications and applications are important and
revealing.
In this respect, the Shirkov-Solovtsov approach may provide a convenient
tool for improving theoretical predictions for exclusive observables,
based on perturbation theory.

\bigskip

\ac
We have benefitted from discussions with Alexander Bakulev, Thorsten
Feldmann, Peter Kroll, Maxim Polyakov, and Christian Weiss.
The work of H.-Ch.K. was supported in part by the Research Institute for
Basic Sciences, Pusan National University under Grant RIBS-PNU-98-203.

\end{document}